\documentclass[aps,twocolumn,showpacs, showkeys,nofootinbib,floatfix]{revtex4-1}

\usepackage{amsmath}
\usepackage{amsfonts}
\usepackage{txfonts}
\usepackage{graphicx}
\usepackage{dcolumn}
\usepackage{bbm}
\usepackage{amssymb}
\usepackage{latexsym}
\usepackage{titlesec}
\usepackage[colorlinks=true, linkcolor=red, citecolor=blue]{hyperref}
\usepackage{longtable}

\begin{document}

\title{Confronting Dark Energy Anisotropic Stress}

\author{Baorong Chang}
\email{changbaorong@dlut.edu.cn}

\author{Lixin Xu}
\email{Corresponding author: lxxu@dlut.edu.cn}

\affiliation{Institute of Theoretical Physics, School of Physics \&
Optoelectronic Technology, Dalian University of Technology, Dalian,
116024, P. R. China}

\affiliation{College of Advanced Science \& Technology, 
Dalian University of Technology, Dalian, 116024, P. R. China}

\affiliation{State Key Laboratory of Theoretical Physics, Institute of Theoretical Physics, Chinese Academy of Sciences}

\begin{abstract}
We use the currently available cosmic observations to probe and constrain an imperfect dark energy fluid which is characterized by a constant equation of state $w$ and a constant speed of viscosity $c^2_{vis}$. The model parameter space was scanned by using Markov chain Monte Carlo method. The results show that the speed of viscosity $c^2_{vis}$ was not well constrained when currently available date sets, which include the cosmic microwave background radiation from {\it Planck}2013, type Ia supernovae and baryon acoustic oscillations, are used. But the cosmic data sets favor phantom dark energy with a negative speed of viscosity $c^2_{vis}$ slightly.  
\end{abstract}



\maketitle

\section{Introduction}

In the last decade, the dark energy having a negative pressure was proposed to explain the currently accelerated expansion of our Universe. However, the nature of dark energy is still unknown. The simplest candidate for it is a cosmological constant, although it suffers from the so-called fine tuning and coincidence problems. In general, this dark energy component was taken as a perfect fluid characterized by its equation of state $w$ and speed of sound $c^2_s$, for example the Chevalier-Polarski-Linder parametrization of the dark energy with equation of state $w(a)=w_0+w_a(1-a)$ and speed of sound $c^2_{s}=1$ \cite{ref:CPL}. However in additions to $w$ and $c^2_s$, there is another important characteristic of a general cosmic fluid which is its anisotropic stress $\sigma$. Basically, while $w$ and $c^2_s$ determine, respectively, the background and perturbative pressure of a fluid that is rotationally invariant, the anisotropic stress $\sigma$ quantities variation of the pressure with direction \cite{ref:Koivisto2006}. A nonzero value of the stress measures the difference between the Newtonian potentials in the conformal Newtonian gauge. When the dark energy is dominated at the late time, its properties are important to determine the evolution of our Universe. Therefore, by using the cosmic observations, one can probe the properties of this dark fluid on the contrary. 

The effects on the CMB power spectrum and linear matter power spectrum due to the anisotropic stress of dark energy were studied, see Refs. \cite{ref:Koivisto2006,ref:Mota2007,ref:Calabrese2011,ref:Dossett2013,ref:Sapone2012,ref:Sapone2013} for examples. In Ref. \cite{ref:Koivisto2006}, it was found that when the equation of state of dark energy is in the range $-1\le w<0$ the increasing anisotropic stress causes a swifter decay of dark energy overdensities and amplification of the integrated Sachs-Wolfe (ISW) effect. The opposite occurs in the case of phantom dark energy ($w<-1$). In Ref. \cite{ref:Mota2007}, the speed of sound and viscosity parameters were constrained by using Type Ia supernovae, large scale structues and WMAP3 CMB data. The authors of Ref. \cite{ref:Calabrese2011} investigated the future constraints on early dark energy achievable by the Planck and CMBPol experiments with anisotropic stress. It was shown that the presence of anisotropic stress can substantially undermine the determination of the early dark energy sound speed parameter. The firs release of {\it Planck} improves the quality of cosmological data extraordinary \cite{ref:Planck}. It allows us to give a tighter constraint to the cosmological parameter space. And the measurements of the redshift space distortion (RSD) can constrain the evolution of the matter perturbations. Therefore, in this brief report, we try to use the currently available cosmic observations, which include the SN Ia, BAO and the first released {\it Planck} CMB data and RSD, to probe and constrain the living space of the sound speed of viscosity $c^2_{vis}$ for a dark energy model of constant equation of state $w$, where the effective speed of sound $c^2_{s,eff}$ is fixed to $1$.     

This paper is structured as follows. In Sec. \ref{sec:basics}, the basic equations for dark energy with anisotropic stress are presented. Where the evolution equations for the density contrast $\delta$, the velocity divergence $\theta$ and the anisotropic stress $\sigma$ for dark energy are given. In Sec. \ref{sec:results}, the constrained results are shown. Sec. \ref{sec:conclusion} is the conclusion.  

\section{Basic Equations for Dark Energy with Anisotropic Stress}\label{sec:basics}

For a general fluid, the energy momentum tensor is defined as
\begin{equation}
T_{\mu\nu}=\rho u_{\mu}u_{\nu}+p h_{\mu\nu}+\Sigma_{\mu\nu},
\end{equation}
where $\rho$ is the energy density and $p$ is the pressure of the fluid; and $u_{\mu}$ is the four-velocity of the fluid, $h_{\mu\nu}\equiv g_{\mu\nu}+u_{\mu}u_{\nu}$ is the projection tensor; $\Sigma_{\mu\nu}$ describes the only spatial inhomogeneity. For a perfect fluid, it is zero. In a isotropic and homogenous universe, $\Sigma_{\mu\nu}$ is also zero at the background level. In this case, it denotes the anisotropic perturbation at the first order. For an adiabatic fluid, the relation $p=p(\rho)$ is respected. Then the evolution of its perturbations is described by the adiabatic speed of sound $c^2_{s,a}$
\begin{equation}
c^2_{s,a}\equiv\frac{\dot{p}}{\dot{\rho}}=w-\frac{\dot{w}}{3\mathcal{H}(1+w)},
\end{equation}   
where $w$ is the equation of state of the fluid $w\equiv p/\rho$; and the dot denotes the derivative with respect to the conformal time $\tau$; and $\mathcal{H}=\dot{a}/a$ is the conformal Hubble parameter. In this case, the relation between the perturbations of $\delta p$ and $\delta\rho$ is related by $\delta p=c^2_{s,a}\delta \rho$. However, for a entropic fluid, the pressure might not be a unique function of the energy density $\rho$. Therefore, there would be another degree of freedom to describe the micro-properties of a general fluid. That is the effective speed of sound $c^2_{s,eff}$
\begin{equation}
c^2_{s,eff}\equiv \frac{\delta p}{\delta \rho}|_{rf},
\end{equation}   
which is defined in the comoving frame of the fluid. When the entropic perturbation vanishes, it is the adiabatic one, i.e. $c^2_{s,eff}=c^2_{s,a}$. Therefore, a perfect fluid is completely descried by its equation of state $w$ and its effective speed of sound $c^2_{s,eff}$. 

However, to fully describe a general fluid and its perturbations, another parameter, the anisotropic stress $\sigma$, should also be included even in a isotropic and homogenous Friedmann-Robterson-Walker (FRW) universe, where the anisotropic stress $\sigma$ can be taken as a spatial perturbation. This quantity describes the difference between Newtonian potential and curvature perturbation in the conformal Newtonian gauge.   

In the synchronous gauge, the perturbation equations of density contrast and velocity divergence for the fluid are written as
\begin{eqnarray}
\dot{\delta}&=&-(1+w)(\theta+\frac{\dot{h}}{2})-3\mathcal{H}(\frac{\delta p}{\delta \rho}-w)\delta,\label{eq:continue}\\
\dot{\theta}&=&-\mathcal{H}(1-3c^2_{s,a})+\frac{\delta p/\delta \rho}{1+w}k^{2}\delta-k^{2}\sigma,\label{eq:euler}
\end{eqnarray}
following the notations of Ma and Bertschinger \cite{ref:MB}; where the anisotropic stress relates to $\Sigma_{\mu\nu}$ via the equality $(\rho+p)\sigma\equiv -(\hat{k}_i\hat{k}_j-\delta_{ij}/3)\Sigma^{ij}$. Using the definition of the effective speed of sound, one can rewrite the above equations into
 \begin{widetext}
 \begin{eqnarray}
 \dot{\delta}&=&-(1+w)(\theta+\frac{\dot{h}}{2})+\frac{\dot{w}}{1+w}\delta-3\mathcal{H}(c^2_{s,eff}-c^2_{s,a})\left[\delta+3\mathcal{H}(1+w)\frac{\theta}{k^2}\right],\\
\dot{\theta}&=&-\mathcal{H}(1-3c^2_{s,eff})\theta+\frac{c^2_{s,eff}}{1+w}k^2\delta-k^2\sigma,
 \end{eqnarray}
\end{widetext}
where the anisotropic stress $\sigma$ respects to the evolution equation
\begin{equation}
\dot{\sigma}+3\mathcal{H}\frac{c^2_{s,a}}{w}\sigma=\frac{8}{3}\frac{c^2_{vis}}{1+w}\left(\theta+\frac{\dot{h}}{2}+3\dot{\eta}\right),\label{eq:shear}
\end{equation}
following Hu \cite{ref:Hustress}, here $c^2_{vis}$ is the viscous speed of sound which controls the relationship between velocity/metric shear and the anisotropic stress. For a relativistic fluid, its value is $1/3$. For a general dark energy fluid, it is free model parameter to be determined by cosmic observations. As suggested in Ref. \cite{ref:Huey}, see also in Ref. \cite{ref:Koivisto2006}, the values of $c^2_{vis}/(1+w)$ should remain positive. Therefore, in this paper, we will consider two parts $w<-1$ (Model I) and $-1<w<0$ (Model II) respectively. The adiabatic initial conditions were adopted. Concretely, for the dark energy component, they are 
\begin{equation}
\delta=\frac{3}{4}(1+w)\delta_r,\quad \theta=\theta_r,\quad \sigma=0,\label{eq:nitialeq}
\end{equation}
as that taken in Ref. \cite{ref:Koivisto2006}.  

\section{Data Sets and Constrained Results} \label{sec:results}

To determine the model parameter space, the type Ia supernovae and BAO data sets of are used to fix the background evolution. The CMB data sets are used to fix the initial conditions. And the redshift space distortion data to fix the evolution of matter perturbations. Concretely, the cosmic observations used in this paper include:

(i) The SN data from SNLS3 which consists of $472$ SN calibrated by SiFTO and SALT2, for the details please see \cite{ref:SNLS3}.   

(ii) For the BAO data points as 'standard ruler', we use the measured ratio of $D_V/r_s$, where $r_s$ is the co-moving sound horizon scale at the recombination epoch, $D_V$ is the 'volume distance' which is defined as
\begin{equation}
D_V(z)=[(1+z)^2D^2_A(z)cz/H(z)]^{1/3},
\end{equation}
where $D_A$ is the angular diameter distance. The BAO data include $D_V(0.106) = 456\pm 27$ [Mpc] from 6dF Galaxy Redshift Survey \cite{ref:BAO6dF}; $D_V(0.35)/r_s = 8.88\pm 0.17$ from SDSS DR7 data \cite{ref:BAOsdssdr7}; $D_V(0.57)/r_s = 13.62\pm 0.22$ from BOSS DR9 data \cite{ref:sdssdr9}. Here the BAO measurements from WiggleZ are not included, as they come from the same galaxy sample as $P(k)$ measurement.

(iii) The full information of CMB which include the recently released {\it Planck} data sets which include the high-l TT likelihood ({\it CAMSpec}) up to a maximum multipole number of $l_{max}=2500$ from $l=50$, the low-l TT likelihood ({\it lowl}) up to $l=49$ and the low-l TE, EE, BB likelihood up to $l=32$ from WMAP9, the data sets are available on line \cite{ref:Planckdata}.

(iv) The present Hubble parameter $H_0 = 73.8\pm 2.4$ [$\text{km s}^{-1} \text{Mpc}^{-1}$] from HST \cite{ref:HST} is used.

(v) The ten $f\sigma_8$ data points from the redshift space distortion (RSD) are used, they are summarized as in Table \ref{tab:fsigma8data}.
\begin{center}
\begin{table}[tbh]
\begin{tabular}{cccl}
\hline\hline 
$\sharp$ & z & $f\sigma_8(z)$ & Survey and Refs \\ \hline
$1$ & $0.067$ & $0.42\pm0.06$ & 6dFGRS~(2012) \cite{ref:fsigma85-Reid2012}\\
$2$ & $0.17$ & $0.51\pm0.06$ & 2dFGRS~(2004) \cite{ref:fsigma81-Percival2004}\\
$3$ & $0.22$ & $0.42\pm0.07$ & WiggleZ~(2011) \cite{ref:fsigma82-Blake2011}\\
$4$ & $0.25$ & $0.39\pm0.05$ & SDSS~LRG~(2011) \cite{ref:fsigma83-Samushia2012}\\
$5$ & $0.37$ & $0.43\pm0.04$ & SDSS~LRG~(2011) \cite{ref:fsigma83-Samushia2012}\\
$6$ & $0.41$ & $0.45\pm0.04$ & WiggleZ~(2011) \cite{ref:fsigma82-Blake2011}\\
$7$ & $0.57$ & $0.43\pm0.03$ & BOSS~CMASS~(2012) \cite{ref:fsigma84-Reid2012}\\
$8$ & $0.60$ & $0.43\pm0.04$ & WiggleZ~(2011) \cite{ref:fsigma82-Blake2011}\\
$9$ & $0.78$ & $0.38\pm0.04$ & WiggleZ~(2011) \cite{ref:fsigma82-Blake2011}\\
$10$ & $0.80$ & $0.47\pm0.08$ & VIPERS~(2013) \cite{ref:fsigma86-Torre2013}\\
\hline\hline
\end{tabular}
\caption{The data points of $f\sigma_8(z)$ measured from RSD with the survey references.}
\label{tab:fsigma8data}
\end{table}
\end{center}

To study the effects of $c^2_{vis}$ to the CMB power spectrum, we modified the publicly available package {\bf CAMB} \cite{ref:CAMB} which is included in {\bf CosmoMC} \cite{ref:MCMC} to calculate the anisotropic power spectrum of CMB. We described the modification of {\bf CAMB} in the appendix \ref{sec:app1}. To use the RSD measurements of the growth rate $f\sigma_8(z)$, we added one module to obtain the corresponding likelihood. Instead of solving the differential equation of the growth factor $f$, we derived $f=d\ln\delta/d\ln a$ directly from the evolution of the perturbations. With addition of two extra parameters $w$ and $c^2_{vis}$, the dimension of model parameter space amounts to eight
\begin{equation}
P\equiv\{\omega_{b},\omega_c, \Theta_{S},\tau, w, c^2_{vis}, n_{s},\log[10^{10}A_{s}]\}.
\end{equation}  
Their priors are summarized in Table \ref{tab:results}. Performing a global fitting to the model parameter space, eight chains were run on the {\it Computing Cluster for Cosmos}; the running was stopped when the Gelman \& Rubin $R-1$ parameter $R-1 \sim 0.01$ was arrived; that guarantees the accurate confidence limits. The obtained results are shown in Table \ref{tab:results}.  

\begin{widetext}
\begin{center}
\begin{table}[tbh]
\begin{tabular}{cccccc}
\hline\hline Parameters & Priors & Mean with errors (I) & Best fit (I) & Mean with errors (II) & Best fit (II) \\ \hline
$\Omega_b h^2$ & $[0.005,0.1]$ & $0.0223_{- 0.0002}^{+0.0002}$ & $0.0223$ & $0.0223_{-0.0002}^{+0.0003}$ & $0.0224$\\
$\Omega_c h^2$ & $[0.01,0.99]$ & $0.116_{-0.002}^{+0.002}$ & $0.116$ & $0.115_{-0.002}^{+0.002}$ & $0.115$\\
$100\theta_{MC}$ & $[0.5,10]$ & $1.0416_{-0.0006}^{+0.0006}$ & $1.0417$ & $1.0418_{-0.0006}^{+0.0006}$ & $1.0413$\\
$\tau$ & $[0.01,0.8]$ & $0.08_{-0.01}^{+0.01}$ & $0.09$ & $0.09_{-0.01}^{+0.01}$ & $0.09$\\
$w$ & $[-5,-1]/[-1,0]$ & $-1.04_{-0.01}^{+0.04}$ & $-1.02$ & $-0.972_{-0.028}^{+0.007}$ & $-0.996$\\
$c^2_{vis}$ & $[-10,0]/[0,10]$ & $-$ & $-1.197$ & $-$ & $7.251$\\
$n_s$ & $[0.5,1.5]$ & $0.966_{-0.006}^{+0.006}$ & $0.967$ & $0.969_{-0.006}^{+0.006}$ & $0.967$\\
${\rm{ln}}(10^{10} A_s)$ & $[2.4,4]$ & $3.07_{-0.02}^{+0.02}$ & $3.08$ & $3.08_{-0.02}^{+0.02}$ & $3.08$\\
\hline
$H_0$ & $73.8\pm 2.4$ & $70.0_{-1.1}^{+1.0}$ & $70.0$ & $68.4_{-0.8}^{+0.9}$ & $69.1$\\
$\Omega_\Lambda$ & $...$ & $0.715_{-0.009}^{+0.009}$ & $0.715$ & $0.704_{-0.009}^{+0.0100}$ & $0.710$\\
$\Omega_m$ & $...$ & $0.285_{-0.009}^{+0.009}$ & $0.285$ & $0.296_{-0.010}^{+0.010}$ & $0.290$\\
$\sigma_8$ & $...$ & $0.82_{-0.01}^{+0.01}$ & $0.82$ & $0.80_{-0.01}^{+0.01}$ & $0.81$\\
$z_{re}$ & $...$ & $10.3_{-1.0}^{+1.0}$ & $10.7$ & $10.8_{-1.1}^{+1.0}$ & $11.0$\\
${\rm{Age}}/{\rm{Gyr}}$ & $...$ & $13.75_{-0.04}^{+0.04}$ & $13.75$ & $13.77_{-0.04}^{+0.04}$ & $13.77$\\
\hline
\multicolumn{1}{c}{$-\ln L$} & &  \multicolumn{2}{c}{$5123.13$} & \multicolumn{2}{c}{$5123.89$} \\
\hline\hline
\end{tabular}
\caption{The mean values with $1\sigma$ errors and the best fit values of the model parameters and derived cosmological parameters for Model I and Model II, where the CMB, SNLS3, BAO and RSD $f\sigma_8$ data sets are used. '$-$' denotes the parameter was not well constrained.}\label{tab:results}
\end{table}
\end{center}
\end{widetext}

\begin{center}
\begin{figure}[tbh]
\includegraphics[width=8.6cm]{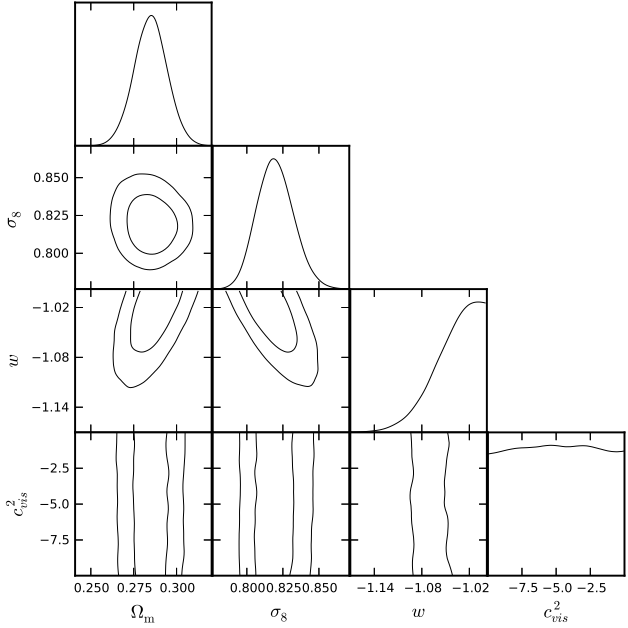}
\caption{The 1D marginalized distribution on individual parameters and 2D contours  with $68\%$ C.L., $95\%$ C.L. and  $99.7\%$ C.L. by using CMB+BAO+SN+RSD data points for Model I.}\label{fig:contourI}
\end{figure}
\end{center}

\begin{center}
\begin{figure}[tbh]
\includegraphics[width=8.6cm]{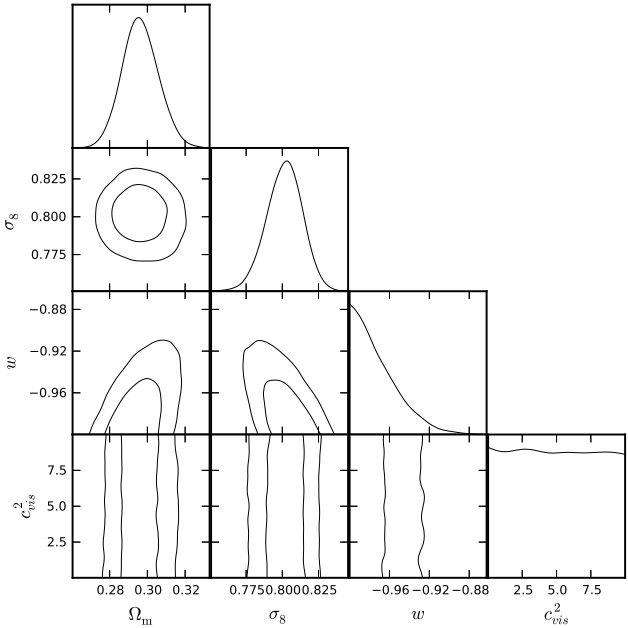}
\caption{The same as Figure \ref{fig:contourI} but for Model II.}\label{fig:contourII}
\end{figure}
\end{center}

\section{Conclusion} \label{sec:conclusion} 

In this brief paper, we have investigated the constraints to the equation of state $w$ and possible anisotropic stress of dark energy by using the currently available cosmic observational data sets which include the CMB of the first 15.5 months from {\it Planck}, SNLS3, SDSS BAO and RSD measurement. To make the evolution of perturbations stable, the values of $c^2_{vis}/(1+w)$ should remain positive. Therefore, we separate the regions of $w$ into two parts: $w<-1$ and $-1<w<0$. Via the Markov chain Mote Carlo method, the model parameter spaces were scanned. In both parts, the currently available cosmic observations cannot give a well constraint. But the difference $\Delta(-\ln L)=0.76$ shows that currently available data sets favor phantom dark energy with a negative speed of viscosity $c^2_{vis}$ slightly. With these observations, one can conclude that currently available cosmic data sets leave large space of the existence of anisotropic stress for dark energy which is characterized by the speed of viscosity $c^2_{vis}$. It is mainly due to the fact that the main constraint to an anisotropic stress of dark energy comes from the ISW effects. However, it is dominated by the cosmic variance.      

\acknowledgements{This work is supported in part by NSFC under the Grants No. 11275035 and "the Fundamental Research Funds for the Central Universities" under the Grants No. DUT13LK01. L. Xu thanks Prof. Z. H. Zhu and Prof. Y. G. Ma for their hospitality during the authorÕs visit in Beijing Normal University.}

\appendix

\section{Modification to the {\bf CAMB} code} \label{sec:app1}

In this appendix section, we give a brief description to the modification of the {\bf CAMB} code, when an anisotropic stress of dark energy is included. At first, the arrays of perturbation evolutions were increased due to the inclusion of an anisotropic stress of dark energy. It was indexed as {\bf w\_ix+2}. In the subroutine {\bf derivs}, under the equations which describe the evolutions of dark energy perturbations, an extra equation (\ref{eq:shear}) was added to evolve the anisotropic stress of dark energy. In the Newtonian gauge, the perturbed line element is
\begin{equation}
ds^2=a^2(\tau)\left[-(1+2\Psi)d\tau^2+(1-2\Phi)\delta_{ij}dx^idx^j\right].
\end{equation} 
When the anisotropic stress of dark energy is nonzero, the summation of potentials become
\begin{equation}
k^2(\Phi+\Psi)=-8\pi G a^2\sum_i\rho_i\Delta_i-8\pi G a^2\sum_i p_i\Pi_i,
\end{equation}   
where $\Delta_i=\delta_i+3\mathcal{H}(1+w_i)\theta_i/k^2$ is the gauge invariant density contrast, and $\Pi_i$ relates to the anisotropic stress $\sigma_i$ via $\sigma_i=\frac{2}{3}\frac{w_i}{1+w_i}\Pi_i$. Therefore an extra contribution to the ISW effect due to the existence of anisotropic stress of dark energy is
\begin{eqnarray}
-k^2 {\bf ISW}_{stress}&=&8\pi G a^2 \sum_i p_i\dot{\Pi}_i\nonumber\\
&-&\mathcal{H}\left[4\cdot8\pi G a^2\sum_i  p_i\Pi_i+8\pi Ga^2\sum_i (3\rho_i-p_i)\Pi_i\right.\nonumber\\
&&\left.-8\pi G a^2\sum_i\frac{w'_i}{w_i}p_i\Pi_i\right].
\end{eqnarray}
In the subroutine {\bf output}, we added them into {\bf dgpi}, {\bf dgpidot} which are defined as
\begin{eqnarray}
dgpi=8\pi G a^2\sum_i  p_i\Pi_i,\\
dgpidot=8\pi G a^2 \sum_i p_i\dot{\Pi}_i,
\end{eqnarray}
and the contribution due to time variation of EoS of dark energy $-8\pi G a^2\sum_i\frac{w'_i}{w_i}p_i\Pi_i$, where $w'_{i}=d w_{i}/d \ln a$.

In the subroutine {\bf initial}, the adiabatic initial conditions for dark energy were also added according to the equations (\ref{eq:nitialeq}).

For the growth function, in the subroutine	{\bf outtransf}, we output and stored the values of $f=d\ln \Delta_m/d\ln a$ at different values $a$ and $k$ in a two dimensional table. Then their values can be used freely.


\begin{thebibliography}{*}

\bibitem{ref:CPL} M. Chevallier, D. Polarski, Int. J. Mod. Phys. D, 10, 213(2001); E. V. Linder, Phys. Rev. Lett., 90, 091301(2003).

\bibitem{ref:Koivisto2006} T. Koivisto, D. Mota, Phys. Rev. D 73, 083502(2006).

\bibitem{ref:Mota2007} D. F. Mota, J. R. Kristiansen, T. Koivisto, N. E. Groeneboom, Mon. Not. Roy. Astron. Soc. 382, 793(2007). 

\bibitem{ref:Calabrese2011} E. Calabrese, R. de Putter, D. Huterer, E. V. Linder, A. Melchiorri, Phys. Rev. D 83, 023011(2011), arXiv:1010.5612 [astro-ph.CO].  

\bibitem{ref:Dossett2013} J. Dossett, M. Ishak, Phys. Rev. D88, 103008(2013), arXiv:1311.0726 [astro-ph.CO].

 \bibitem{ref:Sapone2012} D. Sapone, E. Majerotto, Phys. Rev. D 85, 123529(2012), arXiv:1203.2157 [astro-ph.CO].
 
 \bibitem{ref:Sapone2013} D. Sapone, E. Majerotto, M. Kunz, B. Garilli, Phys. Rev. D 88, 043503(2013), arXiv:1305.1942 [astro-ph.CO].   

\bibitem{ref:Planck} P. A. R. Adeetal., Planck Collaboration, arXiv:1303.5062 [astro-ph.CO].

\bibitem{ref:MB} C.-P Ma and E. Bertschinger, Astrophys. J. 455, 7(1995).

\bibitem{ref:Hustress} W. Hu, Astrophys. J. 506, 485(1998).

\bibitem{ref:Huey} G. Huey, astro-ph/0411102.

\bibitem{ref:CAMB} A. Lewis, A. Challinor, A. Lasenby, Astrophys. J. 538, 473(2000); http://camb.info.

\bibitem{ref:MCMC} A. Lewis and S. Bridle, Phys. Rev. D 66, 103511 (2002); http://cosmologist.info/cosmomc/.

\bibitem{ref:SNLS3} J. Guyetal, A \& A 523, A7 (2010); M. Sullivan, et al, ApJ737, 102 (2011).

\bibitem{ref:BAO6dF} F. Beutler, et al., Mon. Not. Roy. Astron. Soc. 416, 3017 (2011) [arXiv:1106.3366 [astro-ph.CO]].

\bibitem{ref:BAOsdssdr7} N. Padmanabhan, et al., Mon. Not. Roy. Astron. Soc. 427, 2132 (2012) [arXiv:1202.0090 [astro-ph.CO]].

\bibitem{ref:sdssdr9} L. Anderson, et al., Mon. Not. Roy. Astron. Soc. 428, 1036 (2013) arXiv:1203.6594 [astro-ph.CO].

\bibitem{ref:Planckdata} http://pla.esac.esa.int/pla/aio/planckProducts.html.

\bibitem{ref:HST} A. G. Riess, et al., ApJ, 730, 119 (2011) [arXiv:1103.2976[astro-ph.CO]].

\bibitem{ref:fsigma81-Percival2004} W.J. Percival \textit{et al.}[The 2dFGRS Collaboration], Mon. Not. Roy. Astron. Soc. \textbf{353}, 1201 (2004).

\bibitem{ref:fsigma82-Blake2011} C. Blake \textit{et al.}, Mon. Not. Roy. Astron. Soc. \textbf{415}, 2876 (2011).

\bibitem{ref:fsigma83-Samushia2012} L. Samushia, W.J. Percival, and A. Raccanelli, Mon. Not. Roy. Astron. Soc. \textbf{420}, 2102 (2012).

\bibitem{ref:fsigma84-Reid2012} B.A. Reid \textit{et al.}, Mon. Not. Roy. Astron. Soc. 426, 2719(2012), [arXiv:1203.6641].

\bibitem{ref:fsigma85-Reid2012} F. Beutler \textit{et al.}, [arXiv:1204.4725].

\bibitem{ref:fsigma8total-Samushia2013} L. Samushia, \textit{et al.}, Mon. Not. Roy. Astron. Soc. \textbf{429}, 1514 (2013), arXiv: 1206.5309 [astro-ph:CO].

\bibitem{ref:fsigma86-Torre2013} S. de la Torre, \textit{et al.}, Astron. Astrophys. 557, A54(2013), [arXiv:1303.2622]. 

\end{thebibliography}
\end{document}